\documentclass[11pt]{article} %
\usepackage{graphicx}
\usepackage{fancyhdr}
\usepackage{makeidx}
\usepackage{amssymb,amsmath}
\usepackage{physics}
\usepackage{upquote}
\usepackage{xcolor}
\usepackage{enumerate}
\newcommand{\bd}{\begin{document}}
	\newcommand{\ed}{\end{document}}
\newcommand{\bc}{\begin{center}}
	\newcommand{\ec}{\end{center}}
\newcommand{\vs}{\vspace}
\newcommand{\hs}{\hspace}
\newcommand{\beq}{\begin{equation}}
\newcommand{\eeq}{\end{equation}}
\newcommand{\beqs}{\begin{eqn*}}
	\newcommand{\eeqs}{\end{eqn*}}
\newcommand{\bq}{\begin{quote}}
	\newcommand{\eq}{\end{quote}}
\newcommand{\lb}{\linebreak}
\newcommand{\mb}{\makebox}
\newcommand{\fb}{\framebox}
\newcommand{\mc}{\multicolumn}
\newcommand{\ben}{\begin{enumerate}}
	\newcommand{\een}{\end{enumerate}}
\newcommand{\bit}{\begin{itemize}}
	\newcommand{\eit}{\end{itemize}}
\newcommand{\ov}{\overline}
\newcommand{\un}{\underline}
\newcommand{\lt}{\left}
\newcommand{\rt}{\right}
\newcommand{\ba}{\begin{array}}
	\newcommand{\ea}{\end{array}}
\newcommand{\beqa}{\begin{eqnarray}}
\newcommand{\eeqa}{\end{eqnarray}}
\newcommand{\beqas}{\begin{eqnarray*}}
	\newcommand{\eeqas}{\end{eqnarray*}}
\newcommand{\bfg}{\begin{figure}}
	\newcommand{\efg}{\end{figure}}
\newcommand{\pad}{\partial}
\newcommand{\nn}{\nonumber}
\newcommand{\la}{\leftarrow}
\newcommand{\ra}{\rightarrow}
\newcommand{\lgla}{\longleftarrow}
\newcommand{\lgra}{\longrightarrow}
\newcommand{\La}{\Leftarrow}
\newcommand{\Ra}{\Rightarrow}
\newcommand{\Lra}{\Leftrightarrow}
\newcommand{\Lgla}{\Longleftarrow}
\newcommand{\Lgra}{\Longrightarrow}
\renewcommand{\a}{\alpha}
\renewcommand{\b}{\beta}
\newcommand{\g}{\gamma}
\newcommand{\G}{\Gamma}
\renewcommand{\d}{\delta}
\newcommand{\D}{\Delta}
\newcommand{\e}{\epsilon}
\newcommand{\eps}{\epsilon}
\newcommand{\s}{\sigma}
\renewcommand{\l}{\lamda}
\newcommand{\m}{\mu}
\newcommand{\n}{\nu}
\renewcommand{\S}{\Sigma}
\newcommand{\p}{\pi}
\newcommand{\om}{\omega}
\newcommand{\Om}{\Omega}
\newcommand{\tri}{\triangle}
\newcommand{\ti}{\times}
\newcommand{\f}{\frac}
\newcommand{\ds}{\displaystyle}
\newcommand{\bm}[1]{\mb{{\boldmath $#1$}}}
\newcommand{\alter}[2]{\lt\{ \ba{ll}#1 \\ #2 \ea \rt.}
\newcommand{\alt}[4]{\lt\{ \ba{ll}#1 & \mb{if \, \,}#2 \\ #3 & \mb{}#4 \ea
	\rt.}
\newcommand{\altn}[4]{\lt\{ \ba{rl}#1 & \mb{if \, \,}#2 \\ #3 & \mb{}#4 \ea
	\rt.}
\newcommand{\altif}[4]{\lt\{ \ba{ll}#1 & \mb{if \, \,}#2 \\ #3 &
	\mb{if \, \,}#4 \ea \rt.}
\newcommand{\altnif}[4]{\lt\{ \ba{rl}#1 & \mb{if \, \,}#2 \\ #3 &
	\mb{if \, \,}#4 \ea \rt.}
\newcounter{algc}
\newcounter{romc}
\newcounter{Alphc}
\newcommand{\bl}{\begin{list}{{\it Step} ~\arabic{algc}~:} {\usecounter{algc}
			\setlength{\topsep}{0pt} \setlength{\itemsep}{0pt}}}
	\newcommand{\el}{\end{list}}
\newcommand{\blr}{\begin{list}{~\roman{romc}~:} {\usecounter{romc}
			\setlength{\topsep}{0pt} \setlength{\itemsep}{0pt}}}
	\newcommand{\elr}{\end{list}}
\newcommand{\bla}{\begin{list}{~\Alph{Alphc}~:} {\usecounter{Alphc}
			\setlength{\topsep}{0pt} \setlength{\itemsep}{0pt}}}
	\newcommand{\ela}{\end{list}}
\newcommand{\tsup}{\textsuperscript}
\newcommand{\tsub}{\textsubscript}

\newtheorem{theorem}{Theorem}
\begin{document}
\title{Inter-layer charge transport controlled by exciton-trion coherent coupling}
\author{Sangeeth Kallatt$^{||}$, Sarthak Das$^{||}$, Suman Chatterjee, and Kausik Majumdar$^*$\\
Department of Electrical Communication Engineering, \\Indian Institute of Science, Bangalore 560012, India\\
$^{||}$These authors contributed equally,\\
$^*$Corresponding author, email: kausikm@iisc.ac.in}
\date{}
\maketitle
{\abstract The possibility of electrical manipulation and detection of charged exciton (trion) before its radiative recombination makes it promising for excitonic devices. Using a few-layer graphene/monolayer WS\tsub{2}/monolayer graphene vertical heterojunction, we report inter-layer charge transport from top few-layer graphene to bottom monolayer graphene, mediated by coherently formed trion state. This is achieved by using a resonant excitation and varying the sample temperature, the resulting change in the WS\tsub{2} bandgap allows us to scan the excitation around the exciton-trion spectral overlap with high spectral resolution. By correlating the vertical photocurrent and {\it in situ} photoluminescence features at the heterojunction as a function of the spectral position of the excitation, we show that (1) trions are anomalously stable at the junction even up to 463 K due to enhanced doping, and (2) the photocurrent results from the ultra-fast formation of trion through exciton-trion coherent coupling, followed by its fast inter-layer transport. The demonstration of coherent formation, high stabilization, vertical transportation and electrical detection of trions marks a step towards room temperature trionics.}	
\newpage
\section*{Introduction:}
Excitons are bound pairs of an electron and a hole, and play a crucial role in a variety of optoelectronic devices. The large optical absorption, large carrier effective mass, small dielectric constant, and strong out of plane confinement in monolayer transition metal dichalcogenides (TMDCs, for example, MX\tsub{2} where M = Mo, W; X = S, Se) lead to the observation of a variety of excitonic quasiparticles (namely, exciton, charged exciton or trion, biexciton) that remain stable even up to room temperature \cite{glazov2014_PRB,wu2015_PRB,mak2013nmat,plechinger2016natcom,you2015naturephysics}. This makes these monolayers and their vertical heterostructures an ideal platform for exploring the physics of excitons and their manipulation \cite{unuchek2018room}.
\\
\\
A trion is a charged exciton constituting of two electrons and one hole ($X^{-}$) or two holes and one electron ($X^{+}$), which exhibits a binding energy on the order of 30 meV \cite{mak2013nmat,jones2013naturenano} in monolayer TMDCs - a number that is about an order of magnitude higher than that is observed in III-V semiconductor quantum wells \cite{kheng_PRL1993,astakhov_PRB2002,bracker_PRB2005}. Since the binding energy is higher than $k_BT$ at 300 K, trions in monolayer TMDCs are stable at room temperature. This makes trions particularly interesting for room temperature optoelectronic applications for two primary reasons. First, trions carry a net charge, allowing us to manipulate and detect trions through electrical probing. Second, the radiative lifetime of trion has been reported in a broad range from few ps to tens of ps \cite{wang2016,hao2017,wang2014,godde2016}, but nonetheless, longer than that of exciton \cite{wang2016,robert2016,chow2017,palummo2015}, providing us time for electrical manipulation before it radiatively recombines in a spontaneous manner. However, many of the perceived applications of trions have not yet been demonstrated experimentally. This is because trion being a heavy particle, exhibits poor in-plane mobility, resulting in relatively large  transit time under electric field. This leads to radiative recombination of trion before collection at the electrodes, leaving little hope for efficient manipulation and detection in lithography limited planar structures.
\\
\\
Vertical layered heterostructures, where monolayer TMDC is encapsulated by a bottom electrode and an optically transparent top electrode provide an ideal solution to this problem. The generated trion in the ultra-thin TMDC sandwiched layer can be swiftly transferred and collected at the vertical electrodes before radiative recombination, and thus can generate a detectable electrical current. However, two important issues must be addressed for successful implementation of this scheme. First, interlayer carrier transfer is an extremely fast process (sub-picosecond timescale \cite{zheng2018,wang2016ncomm,ceballos2014_ACSnano,hong2014_Naturenano}). Any incoherent trion formation process (for example, through phonon scattering following non-resonant excitation) will be inefficient at the heterojunction, as photo-carriers will be transferred to the electrodes before formation of trion. This calls for the need of an ultra-fast trion formation process, for example, through coherent exciton-trion coupling \cite{hao2017,singh_PRL2014,hao_NL2016,Singh_PRB2016,shepard_ACSNano2017} following a resonant excitation. Second, due to strong screening by the adjacent vertical electrodes, the binding energy of the trion is likely to be suppressed \cite{gupta2017direct}, which can adversely affect its stability at room temperature.
\\
\\
In this work, exploiting efficient coherent coupling between exciton and trion driven by resonant excitation, we demonstrate strong photocurrent in a few-layer graphene (FLG)/monolayer WS\tsub{2}/monolayer graphene (MLG) vertical junction. The photocurrent results from electron hopping from top FLG to bottom MLG mediated through the coherently formed trion state in the WS\tsub{2} sandwiched layer. This is achieved by scanning the spectral overlap between the exciton and the trion using a fixed wavelength excitation at varying temperature. We further demonstrate an anomalous stabilization of trion at the heterojunction at elevated temperatures (up to 463 K) arising due to enhanced doping in the WS\tsub{2} sandwiched layer.
\\
\\
\section*{Results and Discussions:}
\begin{figure}[!hbt]
		\centering
		\includegraphics[scale=0.5]{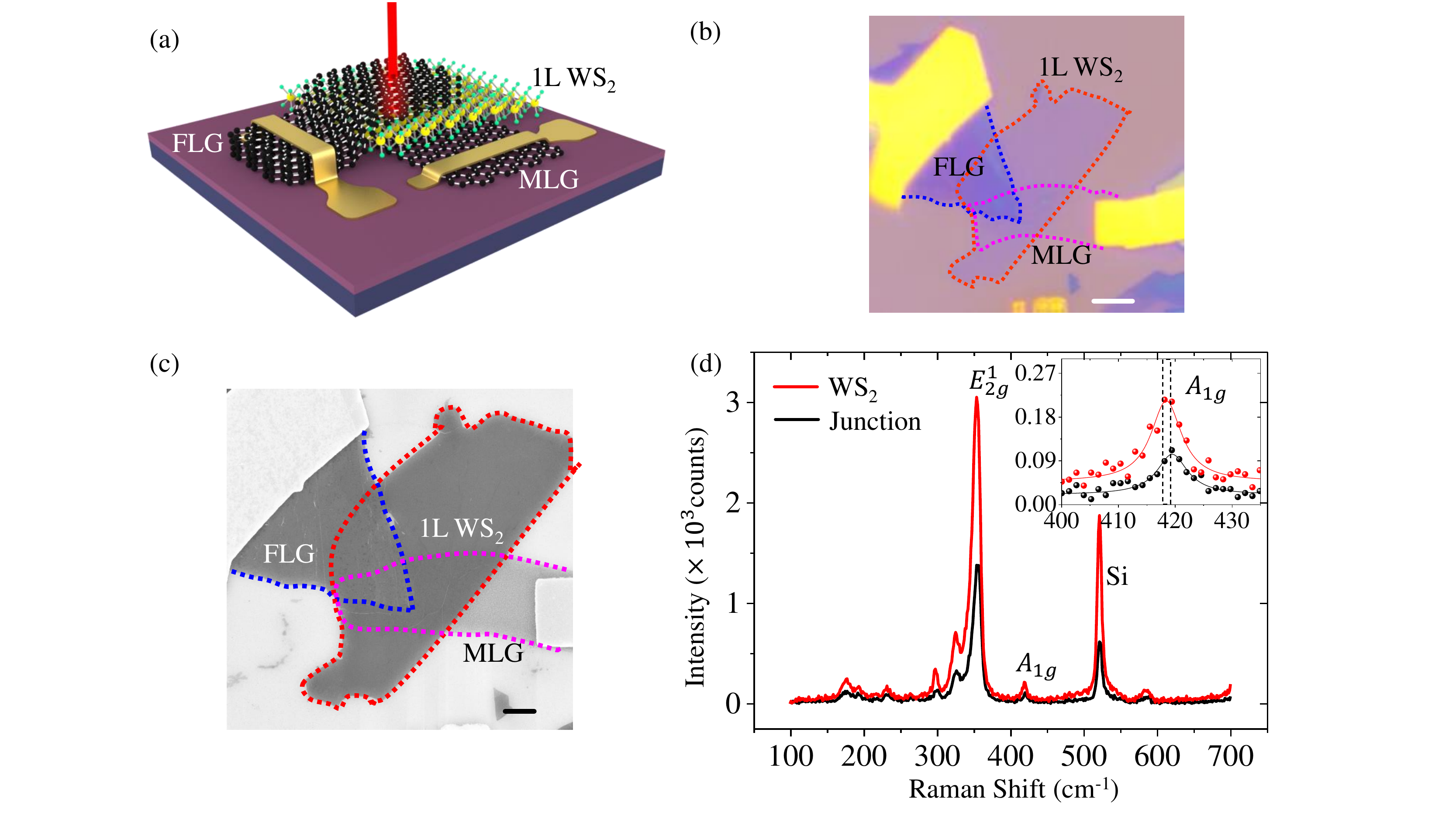}
		\caption{\textbf{FLG/monolayer WS\tsub{2}/MLG vertical heterojunction.} (a) Schematic of the vertical device using monolayer WS\tsub2 sandwiched between few-layer graphene (FLG) on top and monolayer graphene (MLG) at the bottom. (b) Optical image of the device. Scale bar is 4 $\mu$m. (c) Scanning electron micrograph of the top view of the device. The dotted lines indicate the individual layers. Scale bar is 2 $\mu$m. (d) The Raman intensity for both the isolated WS\tsub2 and the junction taken at 296 K with 532 nm laser excitation. Inset: Hardening of the A\tsub1g peak on the junction by about 1 cm\tsup{-1}.}\label{fig:schematic}
	\end{figure}

\begin{figure}[!hbt]
		\centering
		\includegraphics[scale=0.5]{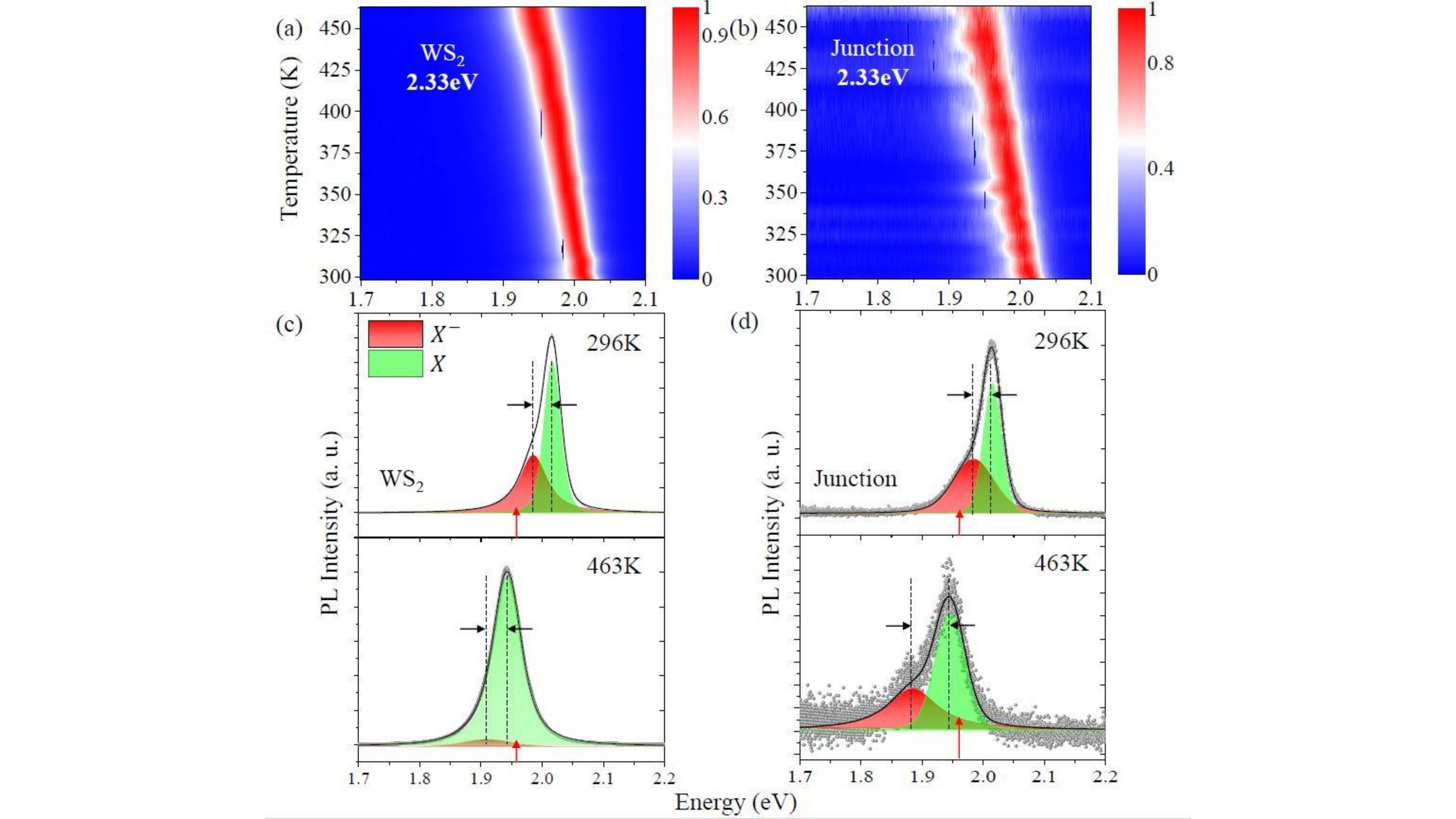}
		\caption{\textbf{Temperature dependent photoluminescence 2.33 eV excitation.} (a-b) Color plot of the normalized PL intensity as a function of temperature ranging from 296 K to 463 K with 2.33 eV laser excitation both on the (a) isolated WS\tsub2 and (b) on the junction. (c-d) PL spectra acquired from (c) the isolated WS\tsub2 and (d) the junction. The top and bottom panels are measured at 296 K and 463 K, respectively.  The data is shown by grey symbols, fitted with voigt function (in black line). The extracted exciton ($X$) and trion ($X^{-}$) peaks shown separately in green and red, respectively. The separation in $X$ and $X^{-}$ peaks is shown by dashed vertical line. The red arrow indicates the excitation position.}\label{fig:532PL}
	\end{figure}

\begin{figure}[!hbt]
		\centering
		\includegraphics[scale=0.5]{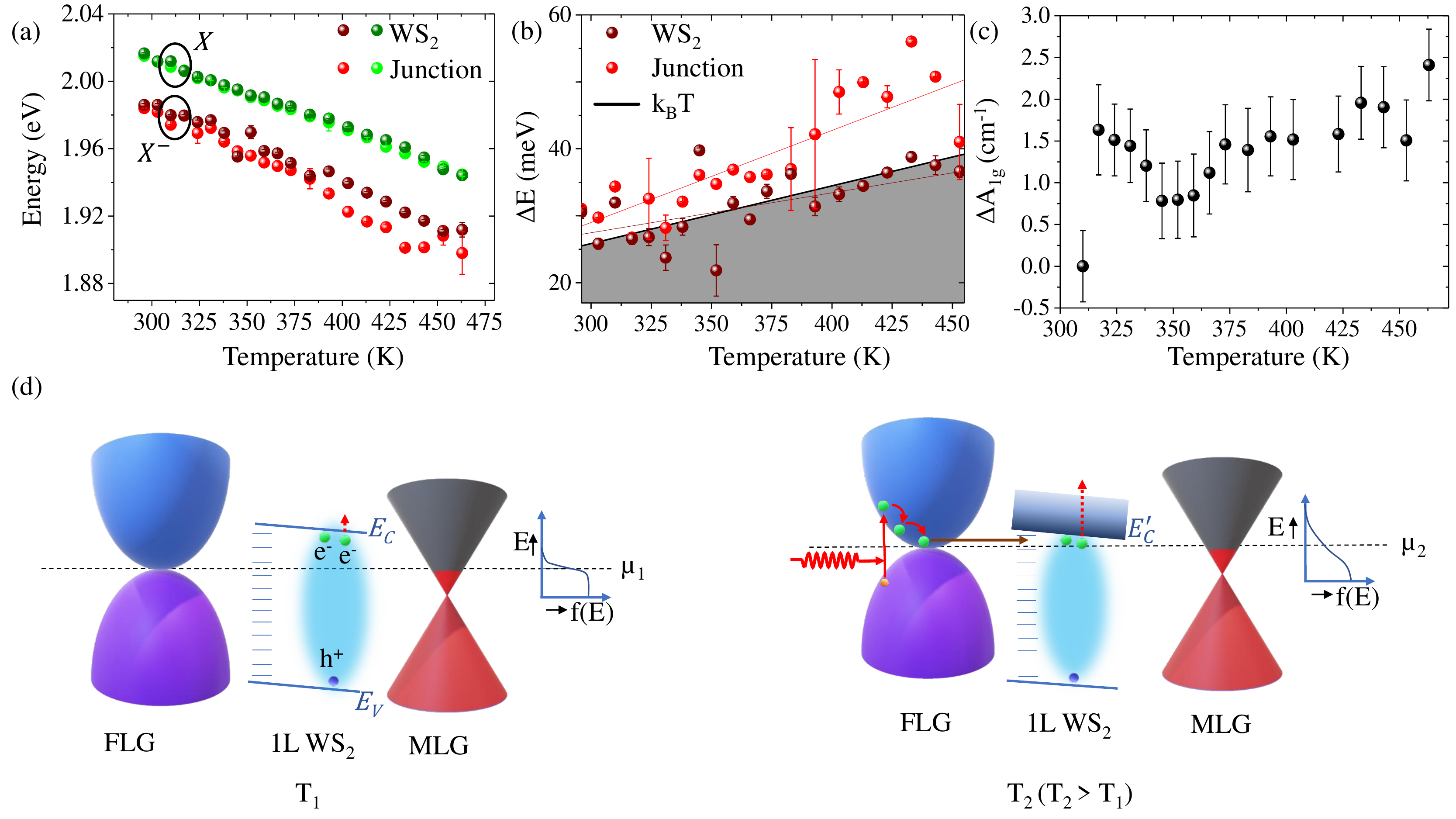}
		\caption{\textbf{Anomalous stabilization of trion at elevated temperature.} (a) The position of the exciton ($X$) and trion ($X^{-}$) peaks for the isolated WS\tsub2 and the junction area as a function of temperature. (b) The separation between $X$ and $X^{-}$ plotted as a function of temperature. The shaded region below the $k_BT$ line is thermally unstable. (c) Relative change in the A\tsub{1g} Raman peak of WS\tsub2 on the junction across the temperature where $\Delta A_{1g} = A_{1g}(T=310 K)-A_{1g}(T)$. (d) Left panel: Band diagram at the heterojunction along the vertical direction at lower temperature. Right panel: Filling of trap states by photo-carriers from FLG and stronger Fermi tail enhances doping at higher temperature. Additional energy is required to ionize the trion, due to filling of lower energy states in the conduction band.}\label{fig:trion}
	\end{figure}

\begin{figure}[!hbt]
		\centering
		\includegraphics[scale=0.5]{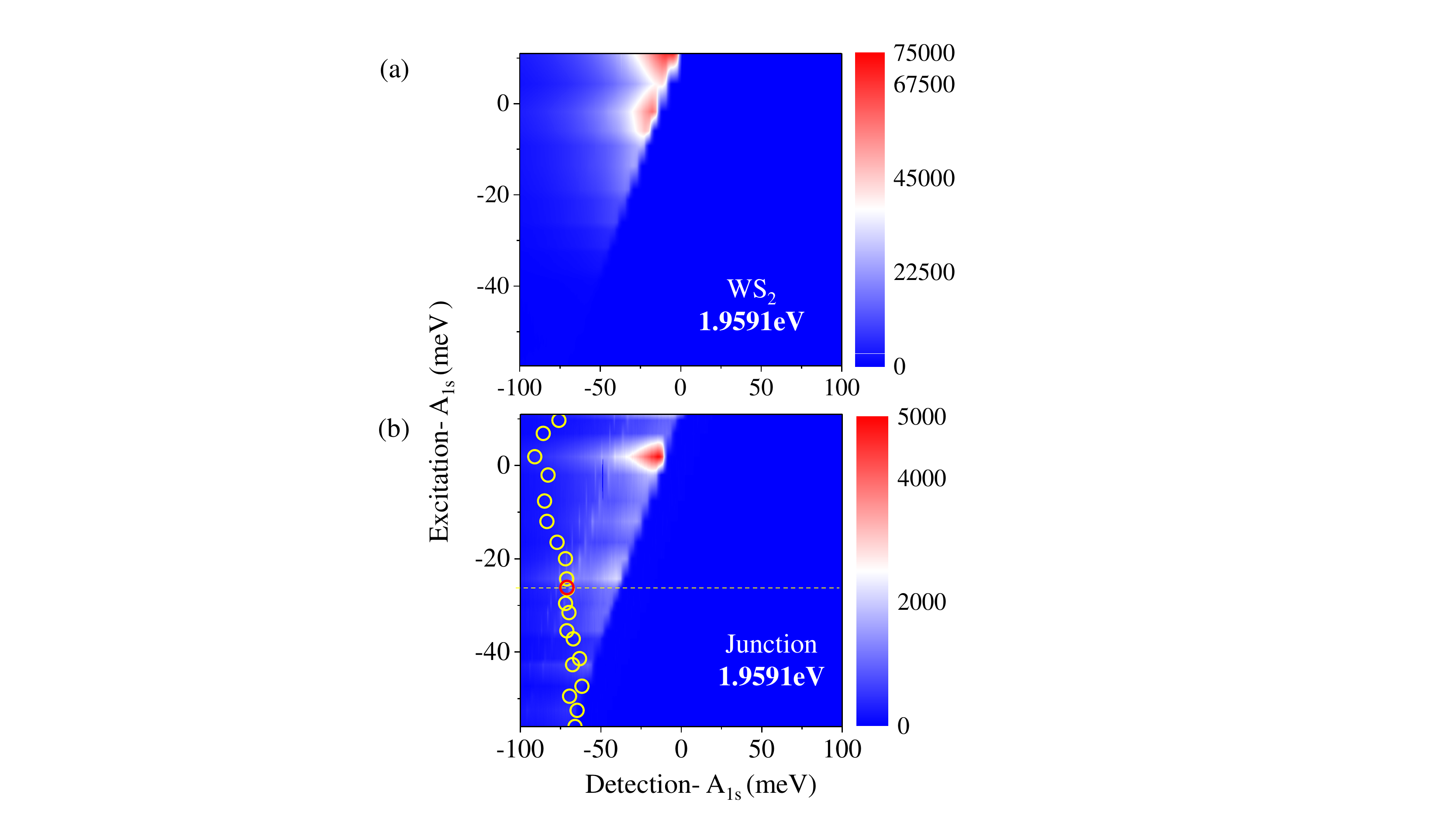}
		\caption{\textbf{Temperature dependent photoluminescence with 1.9591 eV excitation.} Color plot of the PL intensity acquired from (a) the isolated WS\tsub2 portion and (b) the junction, at 1.9591 eV excitation. The vertical and the horizontal axes represent the excitation and the detected photon energy relative to the A\tsub{1s} peak. The sharp Raman peaks are discernable in (b) due to PL quenching \cite{xie_ACS2009}. The open circles represent the energy points 35 meV below the trion peak position, with the peak at the red circle, corresponding to Figure \ref{fig:analysis}b. The horizontal dashed line represents the excitation position that gives maximum photocurrent.}\label{fig:633PL}
	\end{figure}

\begin{figure}[!hbt]
		\centering
		\includegraphics[scale=0.48]{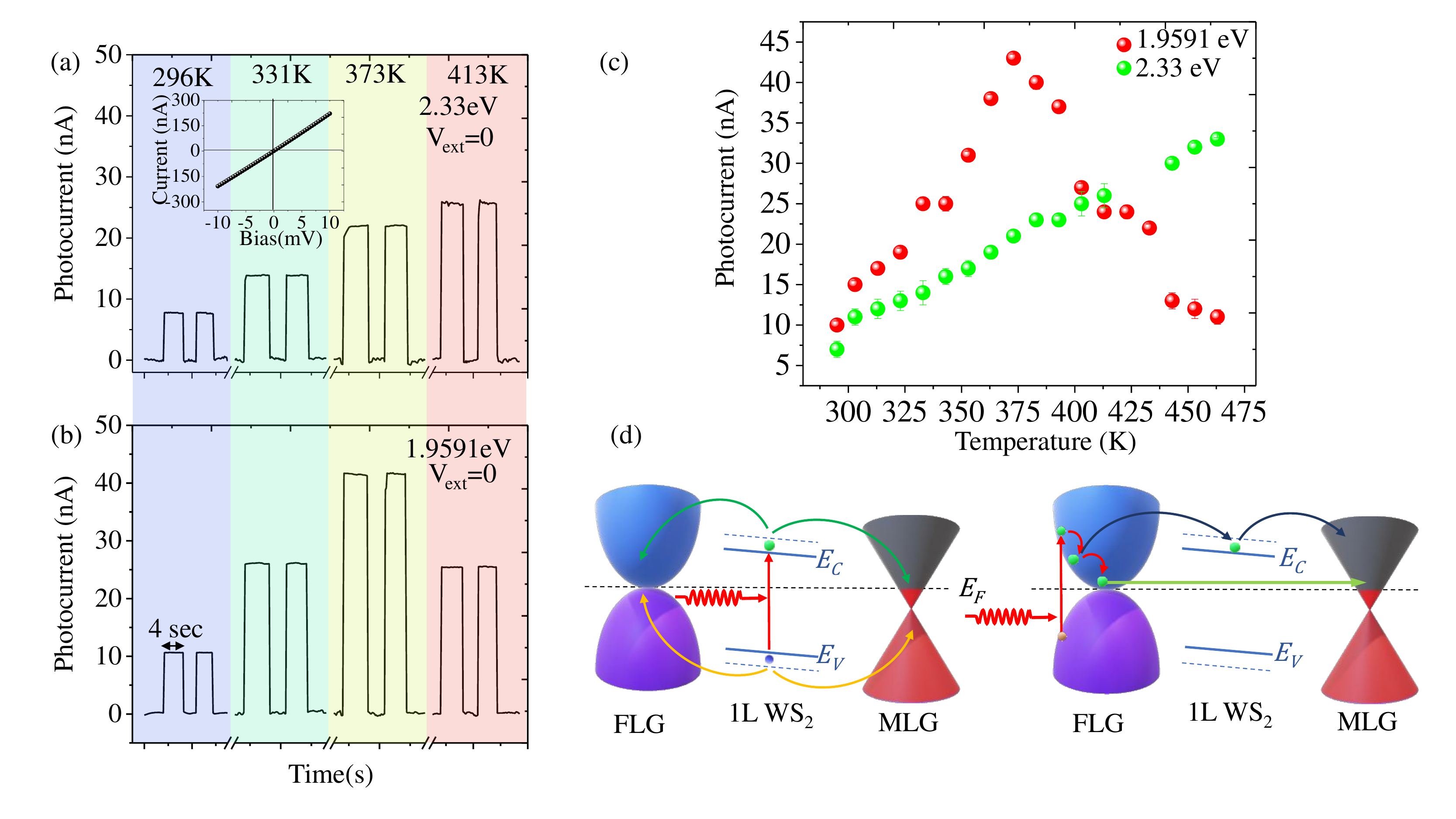}
		\caption{\textbf{Photocurrent from the heterojunction.} (a-b) The transient photocurrent ($I_{ph}$) at zero external bias at four sample temperatures (296 K, 331 K, 373 K and 413 K) with two different excitation energy, (a) 2.33 eV (off-resonant) and (b) 1.9591 eV (resonant). Both rise time and fall time are smaller than $\sim 10$ ms which is the resolution of the measurement setup. Inset of (a): Current-voltage characteristics in the dark at 296 K. (c) Measured photocurrent, as obtained by averaging over several cycles, as a function of sample temperature with 2.33 eV (in green) and 1.9591 eV (in red) excitation. (d) Schematic band diagram representing two photocurrent mechanisms when excited with 2.33 eV photons, namely, WS\tsub2 absorption (left panel), and graphene absorption (right panel).}\label{fig:Iph}
	\end{figure}

\begin{figure}[!hbt]
		\centering
		\includegraphics[scale=0.4]{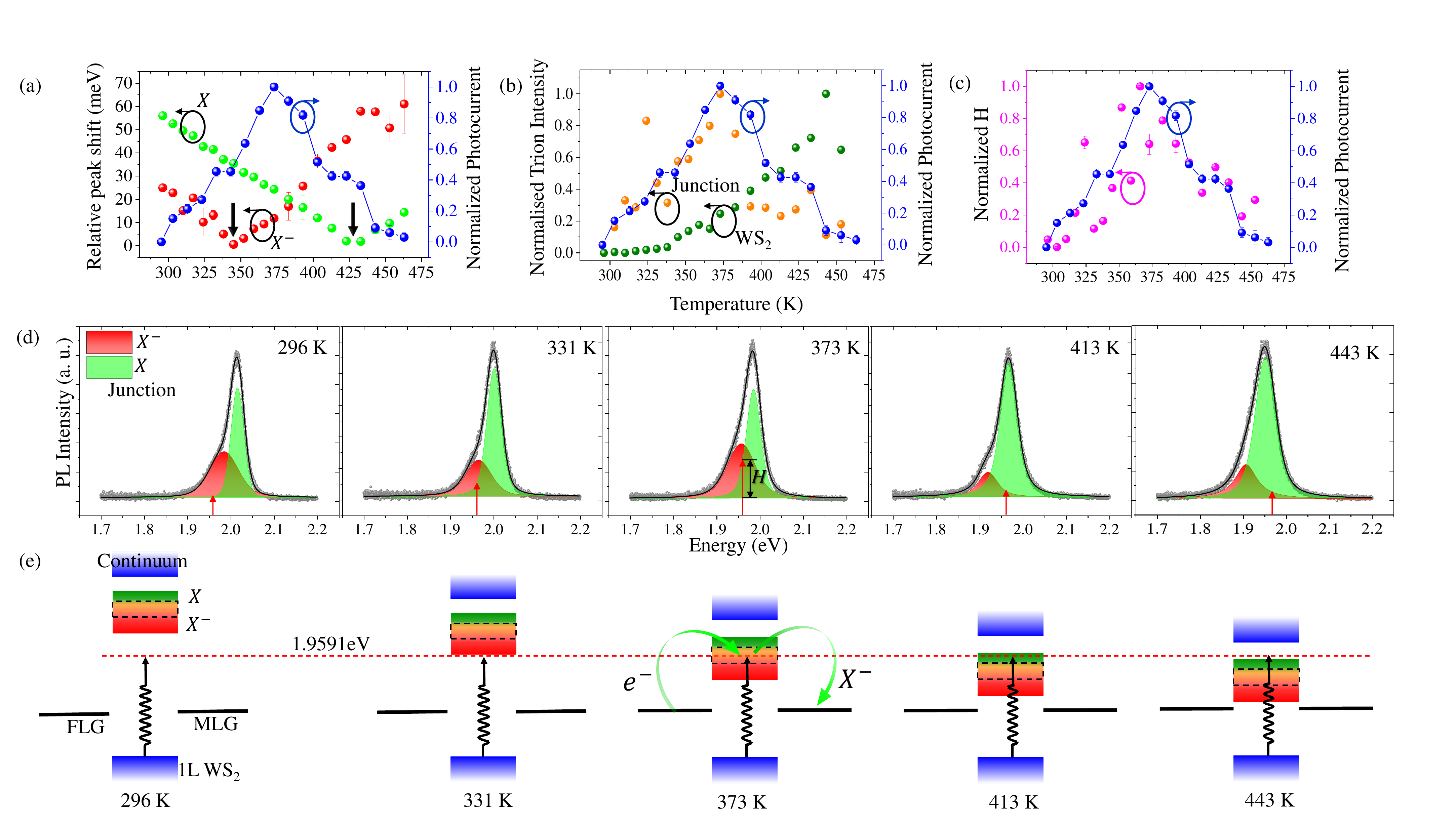}
		\caption{\textbf{Exciton-trion coupling and trion transport probed by resonant photocurrent.} (a) The normalized photocurrent (in blue spheres, same right axis for a-c) as a function of temperature, with 1.9591 eV excitation. The relative position of the 1.9591 eV excitation (left axis) with respect to the $X$ (in green sphere) and $X^{-}$ (in red sphere). The dips, indicated by black arrows correspond to the excitation resonant to the exciton and the trion peaks. (b) Normalized intensity at 35 meV below the trion peak for junction (in orange spheres, left axis) and for isolated WS\tsub2 (in olive spheres) with 1.9591 eV excitation. The corresponding positions are shown by open circles in Figure \ref{fig:633PL}b. (c) Temperature dependent height of the $X$-$X^{-}$ overlap region (magenta spheres on the left axis) at the excitation position (1.9591 eV), normalized by the $X$-$X^{-}$ spectral overlap area. (d) PL spectra at five different temperatures, with fitted $X$ and $X^{-}$ peaks. The excitation position is shown with the red arrow. $H$ is explained in the middle panel. (e) The schematic of the mechanism of the charge transport at the same temperatures as in (e). The excitation is shown by red dotted line. The black dashed rectangle represents the $X$-$X^{-}$ spectral overlap. The middle panel represents peak photocurrent situation. Electron from top FLG hops to the resonantly formed trion state, followed by trion transfer to bottom MLG.}\label{fig:analysis}
	\end{figure}

Figure \ref{fig:schematic}a depicts a schematic diagram of the FLG/monolayer WS\tsub{2}/MLG vertical heterojunction device used in this work. The details of the device fabrication are provided in the Methods section. The optical image and scanning electron micrograph of the fabricated device are shown in Figure \ref{fig:schematic}b-c. Figure \ref{fig:schematic}d shows the Raman spectra obtained using a 532 nm laser at 296 K, both on the isolated WS\tsub{2} portion and on the junction. Compared to the isolated WS\tsub{2}, the Raman signal is slightly suppressed at the junction. Also, the A\tsub{1g} peak at the junction is hardened by about 1 cm\tsup{-1} with respect to the isolated portion (see inset of Figure \ref{fig:schematic}d). This suggests a change in doping in the monolayer WS\tsub{2} at the junction due to depletion of carriers to graphene resulting from band offset.
\\
\\
The fabricated device is mounted on a thermal chuck and the device terminals are connected to Keithley 2636B SMUs through micromanipulators for electrical measurement. At every temperature step, the device is illuminated by a laser beam of photon energy 2.33 eV and 1.9591 eV through a 50X objective, and the photocurrent is measured while acquiring {\it in situ} photoluminescence (PL) spectra. The incident laser power is kept below 8.5 $\mu$W to avoid any undesirable effect due to laser induced heating.
\\
\\
Figure \ref{fig:532PL}a-b illustrate the acquired temperature dependent PL intensity in a color plot for the isolated WS\tsub{2} and the junction, respectively, when excited with photons of energy 2.33 eV. The horizontal axis is the detected photon energy and the vertical axis is the sample temperature varying from 296 K to 463 K. The individual spectra at different temperatures are shown in Supplementary Information S1. The sharp red shift of the A\tsub{1s} exciton ($X$) peak with an increase in temperature is a result of the reduction in quasiparticle bandgap. On the junction, the PL intensity of both the $X$ and the trion ($X^{-}$) peaks is significantly quenched (about 10 times) irrespective of the temperature, as illustrated in Supplementary Information S2. This quenching with 2.33 eV excitation can be attributed to two effects. First, since 2.33 eV excitation is off-resonant to $X$, the formation of the exciton has to happen through the relaxation of energy, by inelastic phonon scattering, which requires a timescale of few picoseconds. Second, once an exciton is formed, on an average it takes on the order of a picosecond for radiative recombination. During these processes, a large fraction of the excitons is transferred to graphene through ultra-fast processes including exciton transfer and non-radiative energy transfer \cite{hill2017_PRB,froehlicher2018_PRX}, quenching the overall PL intensity.
\\
\\
The individual PL spectra acquired from the isolated WS\tsub{2} and the junction are shown in Figure \ref{fig:532PL}c-d at 296 K and 463 K. The obtained PL peaks can be fitted with voigt function to extract the $X$ and $X^{-}$ peaks. Note that, at elevated temperature, the $X^{-}$ spectral strength is almost negligible on isolated WS\tsub{2}, while it is surprisingly strong at the junction. This anomalous trion-stabilization at the junction at higher temperature in spite of screening effect is also supported by the enhanced separation between $X$ and $X^{-}$ peaks in the bottom panel of Figure \ref{fig:532PL}d. The position of the fitted $X$ and $X^{-}$ peaks is systematically plotted as a function of temperature in Figure \ref{fig:trion}a. The corresponding separation between the two peaks ($\Delta E = E_X-E_{X^{-}}$) is shown in Figure \ref{fig:trion}b. We also show the $k_BT$ line indicating the thermal stability region. $\Delta E$ increases with temperature for both isolated WS\tsub{2} and at the junction. The rate of increment of $\Delta E$ with temperature is faster (slower) than the $k_BT$ line at the junction (isolated WS\tsub{2}) portion. This is in agreement with the vanishingly small trion intensity at 463 K at the isolated WS\tsub{2} portion in Figure \ref{fig:532PL}c (bottom panel), while at the junction, we observe a strong trion peak in Figure \ref{fig:532PL}d (bottom panel). This anomalous behaviour at the junction can be attributed to an increasing doping with temperature in the junction. $\Delta E$ is given by \cite{mak2013nmat,stebe_Elsvier1989,huard_PRL2000}
\beq
\Delta E = E_{bT} + \delta E_n
\eeq
where $E_{bT}$ is the trion binding energy and $\delta E_n$ is the additional energy required for trion `ionization' to knock an electron to an empty state in the conduction band. By noting that the A\tsub{1g} Raman peak position of WS\tsub{2} is strongly modulated by electron-phonon coupling \cite{kaasbjerg_PRB2012}, it can be used to monitor the doping effect at the junction. In Figure \ref{fig:trion}c, we plot the shift of the WS\tsub{2} A\tsub{1g} Raman peak in the junction with temperature, showing a softening of $2.5$ cm\tsup{-1}. Note that the increase in temperature would only soften the A\tsub{1g} peak by $1.5$-$1.7$ cm\tsup{-1} \cite{yan_ACSnano2014,sahoo_JPCC2013,kallatt_JPCL2016}, suggesting the additional softening due to electron-phonon coupling arising from increased n-type doping \cite{Biswanath}. The origin of the enhanced n-type doping with temperature at the junction stems from the bandgap defect states in WS\tsub{2}. Such defect states are well studied in TMDC materials, and are known to produce sub-bandgap luminescence peaks at low temperature \cite{plechinger2016natcom,koperski2017Nanophotonics,NaglerPhilipp_PhysRevLett2018,vaclavkova2018_nanotechnology}. A representative PL spectrum of monolayer WS\tsub{2} taken at 3.7 K is shown in Supplementary Information S3. With an increase in temperature, due to a change in the Fermi-Dirac probability, the trap states get more activated. Upon illumination, the photogenerated carriers in the top FLG film populate these trap states, changing the doping in the WS\tsub{2} film, and in turn moving the Fermi level closer to the conduction band edge (see Figure \ref{fig:trion}d). The combined effect of stronger Fermi tail and enhanced doping results in higher probability of filled states at the conduction band edge and statistically it becomes increasingly difficult to ionize the trion by promoting one electron to the conduction band \cite{mak2013nmat,stebe_Elsvier1989,huard_PRL2000}. This in turn results in the anomalous stabilization of the trion at the junction at elevated temperature.
\\
\\
We note from Figure \ref{fig:532PL}a that by changing the sample temperature, the $X$ peak can be shifted from 2.016 eV at 296 K to 1.945 eV at 463 K ($\Delta E=71$ meV). Thus, using a 1.9591 eV excitation at different sample temperatures, we can effectively scan the entire spectral range around $X^{-}$ and $X$, as shown by the red arrows in Figure \ref{fig:532PL}c-d. At $\sim$ 353 K, 1.9591 eV photons resonantly excite the $X^{-}$ peak. With an increase in temperature, around 373 K, the photons with the same energy resonantly excite the spectral overlap region of $X$ and $X^{-}$. With further increase in temperature, around 430 K, $X$ peak is resonantly excited. Figure \ref{fig:633PL}a-b illustrate the PL intensity in a color plot acquired from the isolated WS\tsub{2} portion and the junction for 1.9591 eV excitation. The vertical and the horizontal axes represent the excitation and the detected photon energy relative to the $X$ peak. The individual spectrum at each temperature is shown in Supplementary Information S4, where the higher energy tail is truncated by the edge filter due to near resonant excitation.
\\
\\
We now discuss the effect of such excitation tuning on the photocarrier transport through the junction. The inset of Figure \ref{fig:Iph}a shows the current-voltage ($I$-$V_{ext}$) characteristics of the vertical device under dark condition at 296 K, when biased from -10 mV to 10 mV. The ohmic nature of the junction suggests strong carrier tunneling between top and bottom graphene through the WS\tsub{2} barrier. Such tunneling process can be further assisted by bandgap states due to defects typically present TMDC films. When TMDC based photodetector devices are excited with light, The bandgap states in TMDCs often result in photogating as well as photoconductive gain in the dc photocurrent \cite{Furchi_NL2014}. In order to minimize such effects for understanding the intrinsic photo-carrier transport mechanism, we measure the transient photocurrent, as shown for four representative temperatures in Figure \ref{fig:Iph}a-b using 2.33 eV and 1.9591 eV excitations, all measured at zero $V_{ext}$. The asymmetric doping in the top FLG layer and SiO\tsub{2} substrate supported MLG layer results in such built-in potential. The extremely thin nature of monolayer WS\tsub{2} causes a strong built-in field, allowing us to achieve a strong zero-bias photocurrent. The step rise in the photocurrent ($I_{ph}$) is measured at different temperatures ($T$) in the range of 296 K to 463 K for both the excitation wavelengths, and plotted as a function of $T$ in Figure \ref{fig:Iph}c. The photocurrent is found to be negligible when the laser spot is away from the junction. To confirm that there is no heating induced degradation in the device, we repeat the measurement cycle three times, and the variation in photocurrent from one cycle to the other is negligible.
\\
\\
$I_{ph}$ increases monotonically with $T$ for off-resonant (2.33 eV) excitation, while it is strongly non-monotonic for resonant (1.9591 eV) excitation, suggesting fundamentally different photocurrent mechanisms at play in these two cases. For 2.33 eV excitation, the photocurrent results from both absorption in WS\tsub{2} as well as in graphene. With encapsulation by graphene from top and bottom, the exciton binding energy reduces in WS\tsub{2}, as well as there is a reduction in the continuum level due to bandgap renormalization \cite{ugeda2014naturemat}. Photons with an energy of 2.33 eV is then likely in the WS\tsub{2} continuum level, creating electron-hole pair, followed by separation by the built-in field. However, the sharp band offset at the WS\tsub{2}/graphene interface on both sides of WS\tsub{2} can significantly nullify any net photocurrent out of this process \cite{KrishnaTED}, as illustrated in the left panel in Figure \ref{fig:Iph}d. This suggests that light absorbed by graphene is likely the primary contributor to the photocurrent. The hot carriers generated in the top FLG relax in ultra-fast ($\sim$ picosecond) time scale \cite{Rana_PRB2009,Rana_PRB2007}. However, since inter-layer transfer can also happen in a similar time scale \cite{zheng2018,wang2016ncomm,ceballos2014_ACSnano,hong2014_Naturenano}, a fraction of the photoelectrons in FLG is injected through WS\tsub{2}, aided by the built-in field, and collected in the bottom MLG, producing a photocurrent (right panel of Figure \ref{fig:Iph}d). Both over the WS\tsub{2} barrier (thermionic process) as well as tunneling through WS\tsub{2} are possible for these hot photoelectrons. With an increase in $T$, the n-type doping in WS\tsub{2} enhances as explained earlier. This helps to reduce the effective barrier height for the electrons in FLG to be injected through WS\tsub{2}, in turn resulting in an increment in the photocurrent. Consequently, the photocurrent exhibits a monotonically increasing behaviour with an increase in temperature.
\\
\\
On the other hand, when 1.9591 eV excitation is used, we observe enhanced $I_{ph}$ coupled with a strong non-monotonic behaviour. To get insights, we re-plot $I_{ph}$ (normalized as $\frac{I_{ph}(T)-I_{ph,min}}{I_{ph,max}}$) versus $T$ in Figure \ref{fig:analysis}a-c (in blue). In Figure \ref{fig:analysis}a, we also show the relative position of the excitation with respect to the $X$ and $X^{-}$ peaks. The dips to zero (represented by black arrows) indicate resonance excitation with the $X$ and $X^{-}$ peaks. The data shows that the peak of $I_{ph}$ does not occur when the excitation is in resonance with either the exciton or the trion, rather $I_{ph}$ peaks somewhere in between. On the other hand, the magnitude of the photocurrent strongly correlates with the temperature dependent PL intensity of the trion (in orange spheres), as indicated in Figure \ref{fig:analysis}b - unambiguously proving that the photocurrent stems from the formation of the trion. Note that, due to near-resonant excitation with 1.9591 eV, the $X^{-}$ peak position is cut-off by the edge filter and thus not visible in a large part of the temperature range. In order to plot for the entire temperature range in Figure \ref{fig:analysis}b, we choose a position that is 35 meV below the $X^{-}$ peak, where the $X^{-}$ peak position is already determined by the 2.33 eV excitation in Figure \ref{fig:532PL}. While plotting the temperature dependent trion PL intensity, we normalize as $\frac{PL(T)-PL_{min}}{PL_{max}}$. The corresponding points are shown by open circles in the PL intensity color plot in Figure \ref{fig:633PL}b, with the peak PL intensity of trion (peak $I_{ph}$) marked by red circle (dashed line). As a contrast, the trion intensity monotonically increases with $T$ in isolated WS\tsub{2}, shown by olive spheres in Figure \ref{fig:analysis}b. The spectral position of the excitation is illustrated by the red arrow for five different characteristic temperatures in Figure \ref{fig:analysis}d, where the fitted exciton and trion spectra are separately shown. As shown in Figure \ref{fig:analysis}c, $I_{ph}$ (and hence the $X^{-}$ intensity) also shows excellent correlation with the overlap height $H$ (defined pictorially in the third panel in Figure \ref{fig:analysis}d) between the exciton and the trion at the excitation position. While plotting, $H$ is normalized by the overlap area between $X$ and $X^{-}$. Note that, the peak $I_{ph}$ occurs when the excitation energy is resonant (at 373 K) with the strongest overlap region between the exciton and the trion.
\\
\\
Such a strong tunability of the photocurrent by scanning the exciton-trion spectral overlap region suggests the crucial role of ultra-fast trion formation in the observed photocurrent. Trion, unlike neutral excitons, being a quasiparticle with a net charge, can play an active role in the generation of the photocurrent. However, in order to contribute to the photocurrent in the present vertical heterostructure, trion must be formed at a timescale that is shorter than the ultra-fast inter-layer transport of excitons. It has been shown in the past that trion can be formed through both coherent and incoherent coupling with excitons \cite{hao2017,singh_PRL2014,hao_NL2016,Singh_PRB2016,shepard_ACSNano2017}. The incoherent coupling takes few picoseconds through phonon emission \cite{hao_NL2016,Singh_PRB2016}, which is larger than the inter-layer transfer time \cite{zheng2018,ceballos2014_ACSnano,wang2016ncomm,hong2014_Naturenano}, and is unlikely to play a significant role in the present device. This suggests coherent formation of trion through coulomb interaction between excitons and electrons plays the primary role in the photocurrent generation. On the junction, these electrons can be efficiently supplied by the top FLG. When the incident photons are resonant with the spectral overlap region of the exciton and the trion, such coherent coupling is strongly enhanced. At low temperature, the lower energy side of the exciton peak is often attributed to defect bound excitons \cite{Singh_PRB2016}. However, in our case, the deconvolution of the fitted $X$ peak using a Voigt function suggests that while the inhomogeneous (Gaussian component) broadening is strong and remains a weak function of temperature, the homogeneous (Lorentzian component) broadening increases steadily with temperature due to activated exciton-phonon scattering (see Supplementary Information S5). Thus the lower energy side of the exciton peak can be efficiently contributed from free exciton as well in the present device. The ultra-fast energy transfer between exciton and trion at excitation resonant to the exciton-trion overlap allows an efficient formation of trion before the exciton is transferred to graphene. Once a trion is formed, a fraction of it is quickly transported to the bottom MLG layer, forming a net charge current flow. Thus the trion state acts like an intermediate many-body state in the sandwiched monolayer WS\tsub{2} which mediates the photo-electron transport from the top FLG film to the bottom MLG film, resulting a strong photocurrent. The mechanism is schematically illustrated at different temperature points in Figure \ref{fig:analysis}e. The peak photocurrent corresponds to panel 3 which resonantly excites the exciton-trion spectral overlap at 373 K. As soon as the incident photon energy becomes off-resonant to the exciton-trion spectral overlap region on either side (see panel 1 and 2 for lower energy side, and panel 3 and 4 for higher energy side of the exciton-trion spectral overlap region), this transport mechanism becomes inefficient, suppressing the photocurrent, and resulting in the overall non-monotonicity of the observed photocurrent. Thus, at the peak photocurrent, we achieve an ultra-fast formation of trion through efficient exciton-trion coherent coupling, followed by a vertical transfer and subsequent electrical detection of the trion. The efficiency of the process is controlled by the spectral location of the excitation as the excitation scans through the exciton-trion spectral overlap region.
\\
\\
In summary, we demonstrated a technique by which using an excitation with fixed photon energy and variable sample temperature, we obtain a high resolution spectral scan of the sample. By exploiting this technique, we perform a spectral scan around the exciton-trion overlap region in an asymmetric vertical few-layer graphene/monolayer WS\tsub{2}/monolayer graphene heterojunction to demonstrate a unique inter-layer charge transport mechanism mediated by a coherently formed many-body (trion) state. The efficiency of this transport mechanism can be effectively tuned by scanning the excitation across the exciton-trion spectral overlap. This provides us a knob to control the vertical charge transport by tuning the degree of exciton-trion coupling. This marks a direct probing of coherent coupling between exciton and trion by the electrical detection of trion and would be useful for fast optoelectronic applications. In general, trion being a heavy particle, is not ideally suited for in-plane charge transport due to poor mobility. However, the present work demonstrates the possibility of ultra-fast trion transport through vertical inter-layer transfer. Such ultra-fast inter-layer trion transport can occur before valley and spin depolarization and thus opens up pathways for electrical probing of valley and spin dynamics. Further, the demonstration of enhanced stability of trion at elevated temperatures and its electrical detection is appealing for trion based optoelectronics without the need for any cooling apparatus.
\section*{Methods}
\textbf{Device fabrication:} We prepare a stack of MLG/monolayer WS\tsub{2}/FLG by using dry transfer technique on a highly doped Si substrate coated with 285 nm thick thermally grown oxide. This stack is heated on a hot plate at $80^\circ$ C for 2 minutes for improved adhesion between layers. Devices are fabricated using standard nanofabrication methods. The substrate is spin coated with PMMA 950C3 and baked on a hot plate at $180^\circ$ C for 2 minutes. This is followed by electron beam lithography with an acceleration voltage of 20 KV, an electron beam current of 230 pA, and an electron beam dose of 210 $\mu$Ccm\tsup{-2}. Patterns are developed using MIBK:IPA solution in the ratio 1:3. Later samples are washed in IPA and dried in N\tsub{2} blow. 10 nm Ni /50 nm Au films are deposited by using DC magnetron sputtering at $3\times10^{-3}$ mBar. Metal lift-off is done by dipping the substrate in acetone for 15 minutes, followed by washing in IPA and N\tsub{2} drying. The oxide at the back side of the wafer is etched by dilute HF solution.
\\
\textbf{Photocurrent measurement:} Devices are mounted on a Linkam thermal stage which is connected to a heater. The laser beam (2.33 eV or 1.9591 eV) is focussed through a 50X objective to these devices with a spot size of approximately 2 $\mu$m. The devices are electrically probed using micromanipulators and Keithley 2636B is used as source meter. Temperature of the stage is stepped from 296 K to 463 K. At each temperature, photocurrent and {\it in situ} photoluminescence measurements are carried out. The reported photocurrent values are obtained averaging over several cycles.
\section*{SUPPLEMENTARY INFORMATION}
Supplementary Information is available on (1) temperature dependent PL spectra with 532 nm excitation, (2) temperature dependent exciton and trion PL quenching on junction, (3) low temperature WS\tsub{2} PL spectrum, (4) temperature photoluminescence spectra with 633 nm excitation, and (5) temperature dependent homogeneous and inhomogeneous exciton broadening.
\section*{Acknowledgements}
This work was supported in part by a grant under Indian Space Research Organization (ISRO), by the grants under Ramanujan Fellowship, Early Career Award, and Nano Mission from the Department of Science and Technology (DST), and by a grant from MHRD, MeitY and DST Nano Mission through NNetRA.
\section*{Competing Interests}
The Authors declare no Competing Financial or Non-Financial Interests.
\section*{Author contribution}
K.M. designed the experiment. S.K. fabricated the devices. S.K. and S.D. performed the measurements. All authors contributed to the analysis of the data, and writing of the paper. S.K. and S.D. contributed equally to this work.
\bibliographystyle{naturemag}

\begin{thebibliography}{10}
\expandafter\ifx\csname url\endcsname\relax
  \def\url#1{\texttt{#1}}\fi
\expandafter\ifx\csname urlprefix\endcsname\relax\def\urlprefix{URL }\fi
\providecommand{\bibinfo}[2]{#2}
\providecommand{\eprint}[2][]{\url{#2}}

\bibitem{glazov2014_PRB}
\bibinfo{author}{Glazov, M.} \emph{et~al.}
\newblock \bibinfo{title}{Exciton fine structure and spin decoherence in
  monolayers of transition metal dichalcogenides}.
\newblock \emph{\bibinfo{journal}{Physical Review B}}
  \textbf{\bibinfo{volume}{89}}, \bibinfo{pages}{201302}
  (\bibinfo{year}{2014}).

\bibitem{wu2015_PRB}
\bibinfo{author}{Wu, F.}, \bibinfo{author}{Qu, F.} \&
  \bibinfo{author}{MacDonald, A.}
\newblock \bibinfo{title}{Exciton band structure of monolayer MoS2}.
\newblock \emph{\bibinfo{journal}{Physical Review B}}
  \textbf{\bibinfo{volume}{91}}, \bibinfo{pages}{075310}
  (\bibinfo{year}{2015}).

\bibitem{mak2013nmat}
\bibinfo{author}{Mak, K.~F.} \emph{et~al.}
\newblock \bibinfo{title}{Tightly bound trions in monolayer MoS2}.
\newblock \emph{\bibinfo{journal}{Nature Materials}}
  \textbf{\bibinfo{volume}{12}}, \bibinfo{pages}{207} (\bibinfo{year}{2013}).

\bibitem{plechinger2016natcom}
\bibinfo{author}{Plechinger, G.} \emph{et~al.}
\newblock \bibinfo{title}{Trion fine structure and coupled spin--valley
  dynamics in monolayer tungsten disulfide}.
\newblock \emph{\bibinfo{journal}{Nature Communications}}
  \textbf{\bibinfo{volume}{7}}, \bibinfo{pages}{12715} (\bibinfo{year}{2016}).

\bibitem{you2015naturephysics}
\bibinfo{author}{You, Y.} \emph{et~al.}
\newblock \bibinfo{title}{Observation of biexcitons in monolayer WSe2}.
\newblock \emph{\bibinfo{journal}{Nature Physics}}
  \textbf{\bibinfo{volume}{11}}, \bibinfo{pages}{477} (\bibinfo{year}{2015}).

\bibitem{unuchek2018room}
\bibinfo{author}{Unuchek, D.} \emph{et~al.}
\newblock \bibinfo{title}{Room-temperature electrical control of exciton flux
  in a van der waals heterostructure}.
\newblock \emph{\bibinfo{journal}{Nature}} \textbf{\bibinfo{volume}{560}},
  \bibinfo{pages}{340} (\bibinfo{year}{2018}).

\bibitem{jones2013naturenano}
\bibinfo{author}{Jones, A.~M.} \emph{et~al.}
\newblock \bibinfo{title}{Optical generation of excitonic valley coherence in
  monolayer WSe2}.
\newblock \emph{\bibinfo{journal}{Nature Nanotechnology}}
  \textbf{\bibinfo{volume}{8}}, \bibinfo{pages}{634} (\bibinfo{year}{2013}).

\bibitem{kheng_PRL1993}
\bibinfo{author}{Kheng, K.} \emph{et~al.}
\newblock \bibinfo{title}{Observation of negatively charged excitons X- in
  semiconductor quantum wells}.
\newblock \emph{\bibinfo{journal}{Physical Review Letters}}
  \textbf{\bibinfo{volume}{71}}, \bibinfo{pages}{1752} (\bibinfo{year}{1993}).

\bibitem{astakhov_PRB2002}
\bibinfo{author}{Astakhov, G.} \emph{et~al.}
\newblock \bibinfo{title}{Binding energy of charged excitons in znse-based
  quantum wells}.
\newblock \emph{\bibinfo{journal}{Physical Review B}}
  \textbf{\bibinfo{volume}{65}}, \bibinfo{pages}{165335}
  (\bibinfo{year}{2002}).

\bibitem{bracker_PRB2005}
\bibinfo{author}{Bracker, A.} \emph{et~al.}
\newblock \bibinfo{title}{Binding energies of positive and negative trions:
  From quantum wells to quantum dots}.
\newblock \emph{\bibinfo{journal}{Physical Review B}}
  \textbf{\bibinfo{volume}{72}}, \bibinfo{pages}{035332}
  (\bibinfo{year}{2005}).

\bibitem{wang2016}
\bibinfo{author}{Wang, H.} \emph{et~al.}
\newblock \bibinfo{title}{Radiative lifetimes of excitons and trions in
  monolayers of the metal dichalcogenide MoS2}.
\newblock \emph{\bibinfo{journal}{Physical Review B}}
  \textbf{\bibinfo{volume}{93}}, \bibinfo{pages}{045407}
  (\bibinfo{year}{2016}).

\bibitem{hao2017}
\bibinfo{author}{Hao, K.} \emph{et~al.}
\newblock \bibinfo{title}{Trion valley coherence in monolayer semiconductors}.
\newblock \emph{\bibinfo{journal}{2D Materials}} \textbf{\bibinfo{volume}{4}},
  \bibinfo{pages}{025105} (\bibinfo{year}{2017}).

\bibitem{wang2014}
\bibinfo{author}{Wang, G.} \emph{et~al.}
\newblock \bibinfo{title}{Valley dynamics probed through charged and neutral
  exciton emission in monolayer WSe2}.
\newblock \emph{\bibinfo{journal}{Physical Review B}}
  \textbf{\bibinfo{volume}{90}}, \bibinfo{pages}{075413}
  (\bibinfo{year}{2014}).

\bibitem{godde2016}
\bibinfo{author}{Godde, T.} \emph{et~al.}
\newblock \bibinfo{title}{Exciton and trion dynamics in atomically thin MoSe2
  and WSe2: Effect of localization}.
\newblock \emph{\bibinfo{journal}{Physical Review B}}
  \textbf{\bibinfo{volume}{94}}, \bibinfo{pages}{165301}
  (\bibinfo{year}{2016}).

\bibitem{robert2016}
\bibinfo{author}{Robert, C.} \emph{et~al.}
\newblock \bibinfo{title}{Exciton radiative lifetime in transition metal
  dichalcogenide monolayers}.
\newblock \emph{\bibinfo{journal}{Physical Review B}}
  \textbf{\bibinfo{volume}{93}}, \bibinfo{pages}{205423}
  (\bibinfo{year}{2016}).

\bibitem{chow2017}
\bibinfo{author}{Chow, C.~M.} \emph{et~al.}
\newblock \bibinfo{title}{Phonon-assisted oscillatory exciton dynamics in
  monolayer MoSe2}.
\newblock \emph{\bibinfo{journal}{npj 2D Materials and Applications}}
  \textbf{\bibinfo{volume}{1}}, \bibinfo{pages}{33} (\bibinfo{year}{2017}).

\bibitem{palummo2015}
\bibinfo{author}{Palummo, M.}, \bibinfo{author}{Bernardi, M.} \&
  \bibinfo{author}{Grossman, J.~C.}
\newblock \bibinfo{title}{Exciton radiative lifetimes in two-dimensional
  transition metal dichalcogenides}.
\newblock \emph{\bibinfo{journal}{Nano Letters}} \textbf{\bibinfo{volume}{15}},
  \bibinfo{pages}{2794--2800} (\bibinfo{year}{2015}).

\bibitem{zheng2018}
\bibinfo{author}{Zheng, Q.} \emph{et~al.}
\newblock \bibinfo{title}{Phonon-coupled ultrafast interlayer charge
  oscillation at van der waals heterostructure interfaces}.
\newblock \emph{\bibinfo{journal}{Physical Review B}}
  \textbf{\bibinfo{volume}{97}}, \bibinfo{pages}{205417}
  (\bibinfo{year}{2018}).

\bibitem{wang2016ncomm}
\bibinfo{author}{Wang, H.} \emph{et~al.}
\newblock \bibinfo{title}{The role of collective motion in the ultrafast charge
  transfer in van der waals heterostructures}.
\newblock \emph{\bibinfo{journal}{Nature Communications}}
  \textbf{\bibinfo{volume}{7}}, \bibinfo{pages}{11504} (\bibinfo{year}{2016}).

\bibitem{ceballos2014_ACSnano}
\bibinfo{author}{Ceballos, F.}, \bibinfo{author}{Bellus, M.~Z.},
  \bibinfo{author}{Chiu, H.-Y.} \& \bibinfo{author}{Zhao, H.}
\newblock \bibinfo{title}{Ultrafast charge separation and indirect exciton
  formation in a MoS2--MoSe2 van der waals heterostructure}.
\newblock \emph{\bibinfo{journal}{ACS Nano}} \textbf{\bibinfo{volume}{8}},
  \bibinfo{pages}{12717--12724} (\bibinfo{year}{2014}).

\bibitem{hong2014_Naturenano}
\bibinfo{author}{Hong, X.} \emph{et~al.}
\newblock \bibinfo{title}{Ultrafast charge transfer in atomically thin MoS2/WS2
  heterostructures}.
\newblock \emph{\bibinfo{journal}{Nature Nanotechnology}}
  \textbf{\bibinfo{volume}{9}}, \bibinfo{pages}{682} (\bibinfo{year}{2014}).

\bibitem{singh_PRL2014}
\bibinfo{author}{Singh, A.} \emph{et~al.}
\newblock \bibinfo{title}{Coherent electronic coupling in atomically thin MoSe2}.
\newblock \emph{\bibinfo{journal}{Physical Review Letters}}
  \textbf{\bibinfo{volume}{112}}, \bibinfo{pages}{216804}
  (\bibinfo{year}{2014}).

\bibitem{hao_NL2016}
\bibinfo{author}{Hao, K.} \emph{et~al.}
\newblock \bibinfo{title}{Coherent and incoherent coupling dynamics between
  neutral and charged excitons in monolayer mose2}.
\newblock \emph{\bibinfo{journal}{Nano Letters}} \textbf{\bibinfo{volume}{16}},
  \bibinfo{pages}{5109--5113} (\bibinfo{year}{2016}).

\bibitem{Singh_PRB2016}
\bibinfo{author}{Singh, A.} \emph{et~al.}
\newblock \bibinfo{title}{Trion formation dynamics in monolayer transition
  metal dichalcogenides}.
\newblock \emph{\bibinfo{journal}{Phys. Rev. B}} \textbf{\bibinfo{volume}{93}},
  \bibinfo{pages}{041401} (\bibinfo{year}{2016}).
\newblock \urlprefix\url{https://link.aps.org/doi/10.1103/PhysRevB.93.041401}.

\bibitem{shepard_ACSNano2017}
\bibinfo{author}{Shepard, G.~D.} \emph{et~al.}
\newblock \bibinfo{title}{Trion-species-resolved quantum beats in MoSe2}.
\newblock \emph{\bibinfo{journal}{ACS Nano}} \textbf{\bibinfo{volume}{11}},
  \bibinfo{pages}{11550--11558} (\bibinfo{year}{2017}).

\bibitem{gupta2017direct}
\bibinfo{author}{Gupta, G.}, \bibinfo{author}{Kallatt, S.} \&
  \bibinfo{author}{Majumdar, K.}
\newblock \bibinfo{title}{Direct observation of giant binding energy modulation
  of exciton complexes in monolayer MoSe2}.
\newblock \emph{\bibinfo{journal}{Physical Review B}}
  \textbf{\bibinfo{volume}{96}}, \bibinfo{pages}{081403}
  (\bibinfo{year}{2017}).

\bibitem{hill2017_PRB}
\bibinfo{author}{Hill, H.~M.} \emph{et~al.}
\newblock \bibinfo{title}{Exciton broadening in Ws2/graphene heterostructures}.
\newblock \emph{\bibinfo{journal}{Physical Review B}}
  \textbf{\bibinfo{volume}{96}}, \bibinfo{pages}{205401}
  (\bibinfo{year}{2017}).

\bibitem{froehlicher2018_PRX}
\bibinfo{author}{Froehlicher, G.}, \bibinfo{author}{Lorchat, E.} \&
  \bibinfo{author}{Berciaud, S.}
\newblock \bibinfo{title}{Charge versus energy transfer in atomically thin
  graphene-transition metal dichalcogenide van der waals heterostructures}.
\newblock \emph{\bibinfo{journal}{Physical Review X}}
  \textbf{\bibinfo{volume}{8}}, \bibinfo{pages}{011007} (\bibinfo{year}{2018}).

\bibitem{stebe_Elsvier1989}
\bibinfo{author}{St{\'e}b{\'e}, B.} \& \bibinfo{author}{Ainane, A.}
\newblock \bibinfo{title}{Ground state energy and optical absorption of
  excitonic trions in two dimensional semiconductors}.
\newblock \emph{\bibinfo{journal}{Superlattices and microstructures}}
  \textbf{\bibinfo{volume}{5}}, \bibinfo{pages}{545--548}
  (\bibinfo{year}{1989}).

\bibitem{huard_PRL2000}
\bibinfo{author}{Huard, V.}, \bibinfo{author}{Cox, R.},
  \bibinfo{author}{Saminadayar, K.}, \bibinfo{author}{Arnoult, A.} \&
  \bibinfo{author}{Tatarenko, S.}
\newblock \bibinfo{title}{Bound states in optical absorption of semiconductor
  quantum wells containing a two-dimensional electron gas}.
\newblock \emph{\bibinfo{journal}{Physical Review Letters}}
  \textbf{\bibinfo{volume}{84}}, \bibinfo{pages}{187} (\bibinfo{year}{2000}).

\bibitem{kaasbjerg_PRB2012}
\bibinfo{author}{Kaasbjerg, K.}, \bibinfo{author}{Thygesen, K.~S.} \&
  \bibinfo{author}{Jacobsen, K.~W.}
\newblock \bibinfo{title}{Phonon-limited mobility in n-type single-layer MoS2
  from first principles}.
\newblock \emph{\bibinfo{journal}{Physical Review B}}
  \textbf{\bibinfo{volume}{85}}, \bibinfo{pages}{115317}
  (\bibinfo{year}{2012}).

\bibitem{yan_ACSnano2014}
\bibinfo{author}{Yan, R.} \emph{et~al.}
\newblock \bibinfo{title}{Thermal conductivity of monolayer molybdenum
  disulfide obtained from temperature-dependent raman spectroscopy}.
\newblock \emph{\bibinfo{journal}{ACS Nano}} \textbf{\bibinfo{volume}{8}},
  \bibinfo{pages}{986--993} (\bibinfo{year}{2014}).

\bibitem{sahoo_JPCC2013}
\bibinfo{author}{Sahoo, S.}, \bibinfo{author}{Gaur, A.~P.},
  \bibinfo{author}{Ahmadi, M.}, \bibinfo{author}{Guinel, M. J.-F.} \&
  \bibinfo{author}{Katiyar, R.~S.}
\newblock \bibinfo{title}{Temperature-dependent raman studies and thermal
  conductivity of few-layer mos2}.
\newblock \emph{\bibinfo{journal}{The Journal of Physical Chemistry C}}
  \textbf{\bibinfo{volume}{117}}, \bibinfo{pages}{9042--9047}
  (\bibinfo{year}{2013}).

\bibitem{kallatt_JPCL2016}
\bibinfo{author}{Kallatt, S.}, \bibinfo{author}{Umesh, G.} \&
  \bibinfo{author}{Majumdar, K.}
\newblock \bibinfo{title}{Valley-coherent hot carriers and thermal relaxation
  in monolayer transition metal dichalcogenides}.
\newblock \emph{\bibinfo{journal}{The journal of physical chemistry letters}}
  \textbf{\bibinfo{volume}{7}}, \bibinfo{pages}{2032--2038}
  (\bibinfo{year}{2016}).

\bibitem{Biswanath}
\bibinfo{author}{Chakraborty, B.} \emph{et~al.}
\newblock \bibinfo{title}{Symmetry-dependent phonon renormalization in
  monolayer mos${}_{2}$ transistor}.
\newblock \emph{\bibinfo{journal}{Phys. Rev. B}} \textbf{\bibinfo{volume}{85}},
  \bibinfo{pages}{161403} (\bibinfo{year}{2012}).
\newblock \urlprefix\url{https://link.aps.org/doi/10.1103/PhysRevB.85.161403}.

\bibitem{koperski2017Nanophotonics}
\bibinfo{author}{Koperski, M.} \emph{et~al.}
\newblock \bibinfo{title}{Optical properties of atomically thin transition
  metal dichalcogenides: observations and puzzles}.
\newblock \emph{\bibinfo{journal}{Nanophotonics}} \textbf{\bibinfo{volume}{6}},
  \bibinfo{pages}{1289--1308} (\bibinfo{year}{2017}).

\bibitem{NaglerPhilipp_PhysRevLett2018}
\bibinfo{author}{Nagler, P.} \emph{et~al.}
\newblock \bibinfo{title}{Zeeman splitting and inverted polarization of
  biexciton emission in monolayer ${\mathrm{ws}}_{2}$}.
\newblock \emph{\bibinfo{journal}{Phys. Rev. Lett.}}
  \textbf{\bibinfo{volume}{121}}, \bibinfo{pages}{057402}
  (\bibinfo{year}{2018}).
\newblock
  \urlprefix\url{https://link.aps.org/doi/10.1103/PhysRevLett.121.057402}.

\bibitem{vaclavkova2018_nanotechnology}
\bibinfo{author}{Vaclavkova, D.} \emph{et~al.}
\newblock \bibinfo{title}{Singlet and triplet trions in WS2 monolayer
  encapsulated in hexagonal boron nitride}.
\newblock \emph{\bibinfo{journal}{Nanotechnology}}
  \textbf{\bibinfo{volume}{29}}, \bibinfo{pages}{325705}
  (\bibinfo{year}{2018}).

\bibitem{Furchi_NL2014}
\bibinfo{author}{Furchi, M.~M.}, \bibinfo{author}{Polyushkin, D.~K.},
  \bibinfo{author}{Pospischil, A.} \& \bibinfo{author}{Mueller, T.}
\newblock \bibinfo{title}{Mechanisms of photoconductivity in atomically thin
  MoS2}.
\newblock \emph{\bibinfo{journal}{Nano Letters}} \textbf{\bibinfo{volume}{14}},
  \bibinfo{pages}{6165--6170} (\bibinfo{year}{2014}).

\bibitem{ugeda2014naturemat}
\bibinfo{author}{Ugeda, M.~M.} \emph{et~al.}
\newblock \bibinfo{title}{Giant bandgap renormalization and excitonic effects
  in a monolayer transition metal dichalcogenide semiconductor}.
\newblock \emph{\bibinfo{journal}{Nature Materials}}
  \textbf{\bibinfo{volume}{13}}, \bibinfo{pages}{1091} (\bibinfo{year}{2014}).

\bibitem{KrishnaTED}
\bibinfo{author}{Murali, K.} \& \bibinfo{author}{Majumdar, K.}
\newblock \bibinfo{title}{Self-powered, highly sensitive, high-speed
  photodetection using ITO/WSe2/SnSe2 vertical heterojunction}.
\newblock \emph{\bibinfo{journal}{IEEE Transactions on Electron Devices}}
  \bibinfo{pages}{4141--4148} (\bibinfo{year}{2018}).

\bibitem{Rana_PRB2009}
\bibinfo{author}{Rana, F.} \emph{et~al.}
\newblock \bibinfo{title}{Carrier recombination and generation rates for
  intravalley and intervalley phonon scattering in graphene}.
\newblock \emph{\bibinfo{journal}{Phys. Rev. B}} \textbf{\bibinfo{volume}{79}},
  \bibinfo{pages}{115447} (\bibinfo{year}{2009}).
\newblock \urlprefix\url{https://link.aps.org/doi/10.1103/PhysRevB.79.115447}.

\bibitem{Rana_PRB2007}
\bibinfo{author}{Rana, F.}
\newblock \bibinfo{title}{Electron-hole generation and recombination rates for
  coulomb scattering in graphene}.
\newblock \emph{\bibinfo{journal}{Phys. Rev. B}} \textbf{\bibinfo{volume}{76}},
  \bibinfo{pages}{155431} (\bibinfo{year}{2007}).
\newblock \urlprefix\url{https://link.aps.org/doi/10.1103/PhysRevB.76.155431}.

\bibitem{xie_ACS2009}
\bibinfo{author}{Xie, L.}, \bibinfo{author}{Ling, X.}, \bibinfo{author}{Fang,
  Y.}, \bibinfo{author}{Zhang, J.} \& \bibinfo{author}{Liu, Z.}
\newblock \bibinfo{title}{Graphene as a substrate to suppress fluorescence in
  resonance raman spectroscopy}.
\newblock \emph{\bibinfo{journal}{Journal of the American Chemical Society}}
  \textbf{\bibinfo{volume}{131}}, \bibinfo{pages}{9890--9891}
  (\bibinfo{year}{2009}).

\end{thebibliography}

\newpage
\newpage
\newpage
\newpage
\newpage
\newpage

\newpage
\begin{figure}[h]
\centering
\includegraphics[scale=0.5]{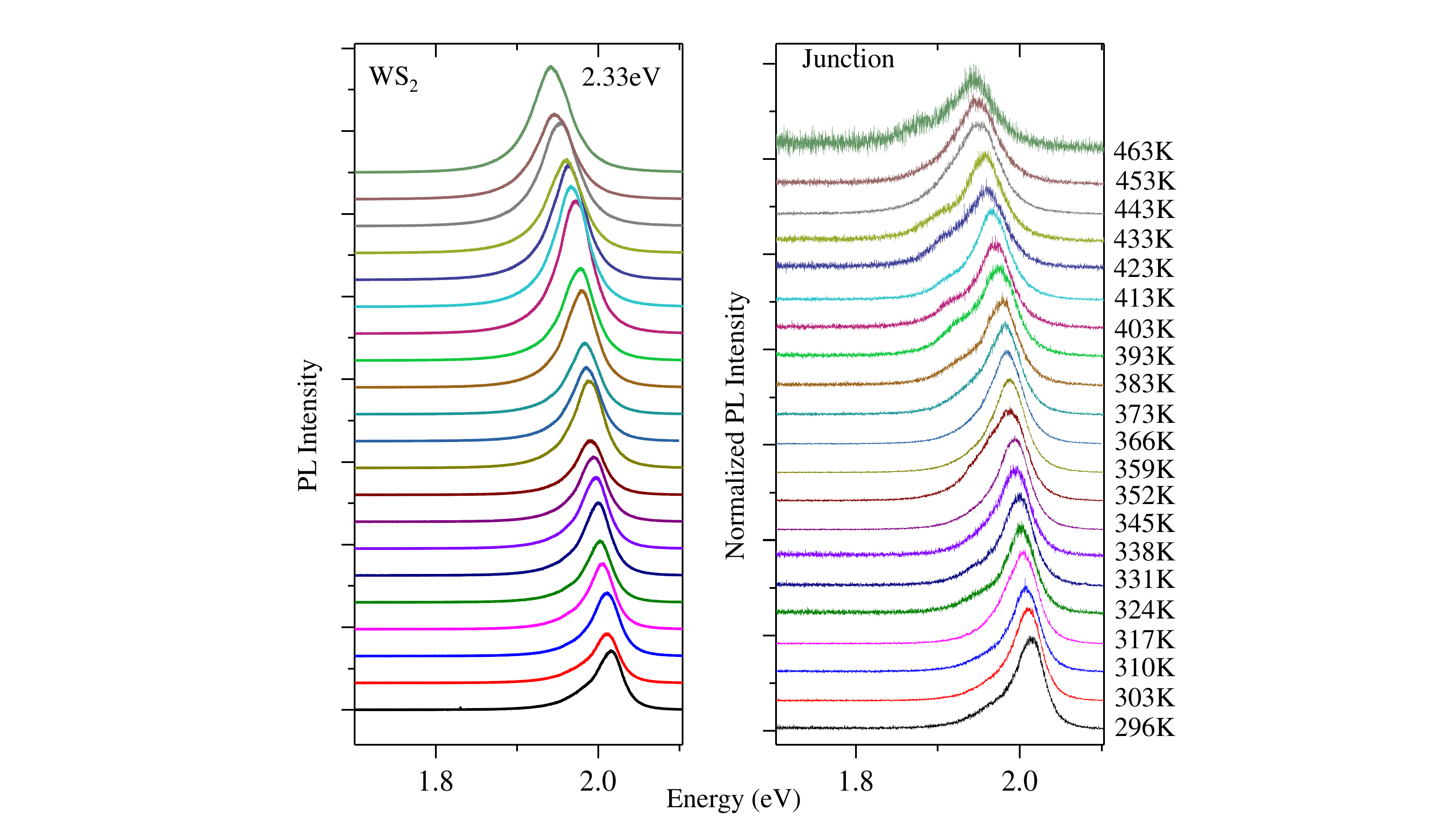}
Figure S1: PL intensity for monolayer WS\tsub2 (left panel) and normalized PL intensity for the junction (right panel) with 2.33 eV laser excitation.
\end{figure}

\newpage
\begin{figure}[hbt]
\centering
\includegraphics[scale=0.5]{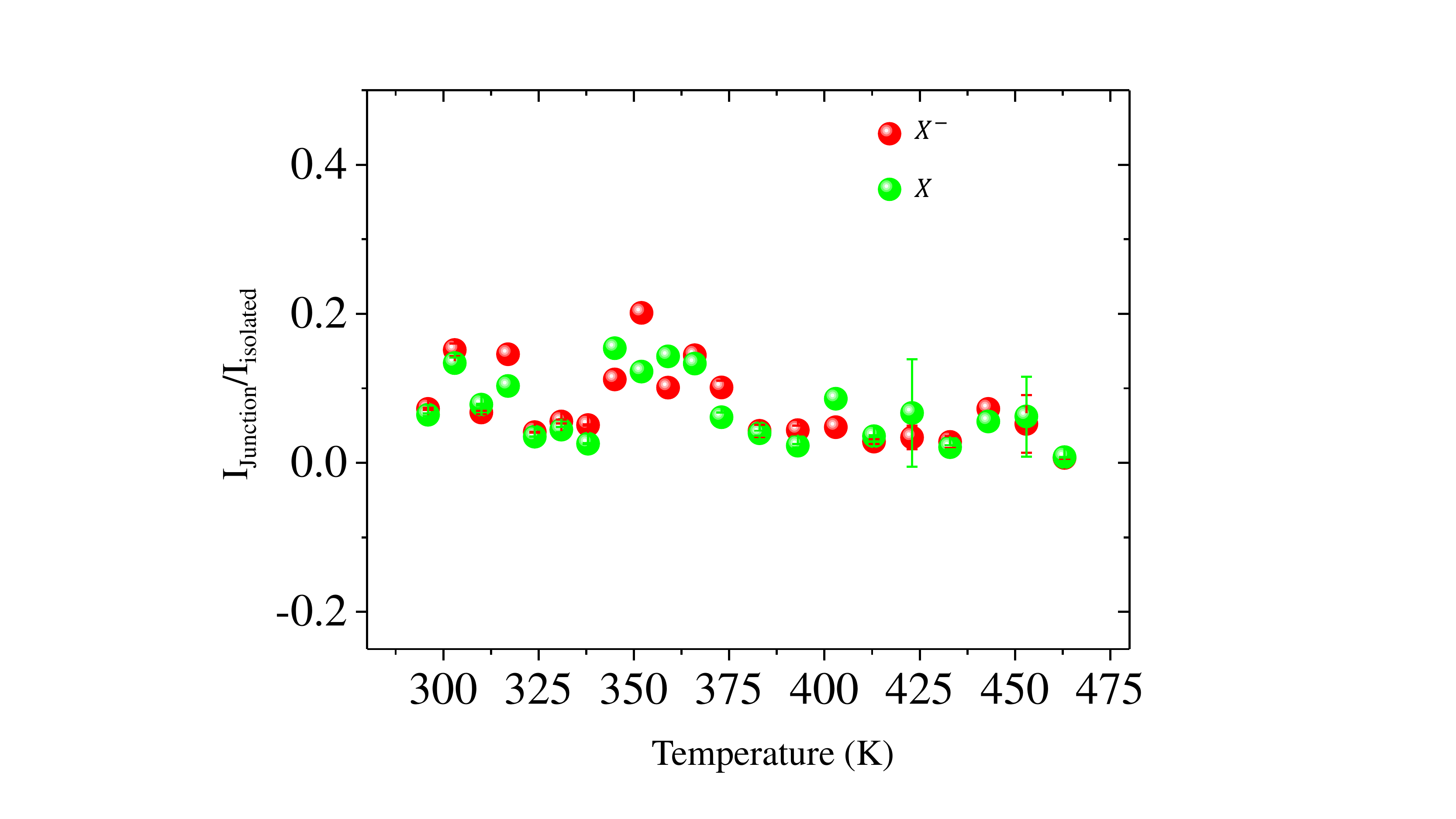}
Figure S2: Quenching ratio (I\tsub{junction}/I\tsub{isolated}) as a function of temperature at the junction for both the exciton and trion.
\end{figure}

\newpage
\begin{figure}[hbt]
\centering
\includegraphics[scale=0.5]{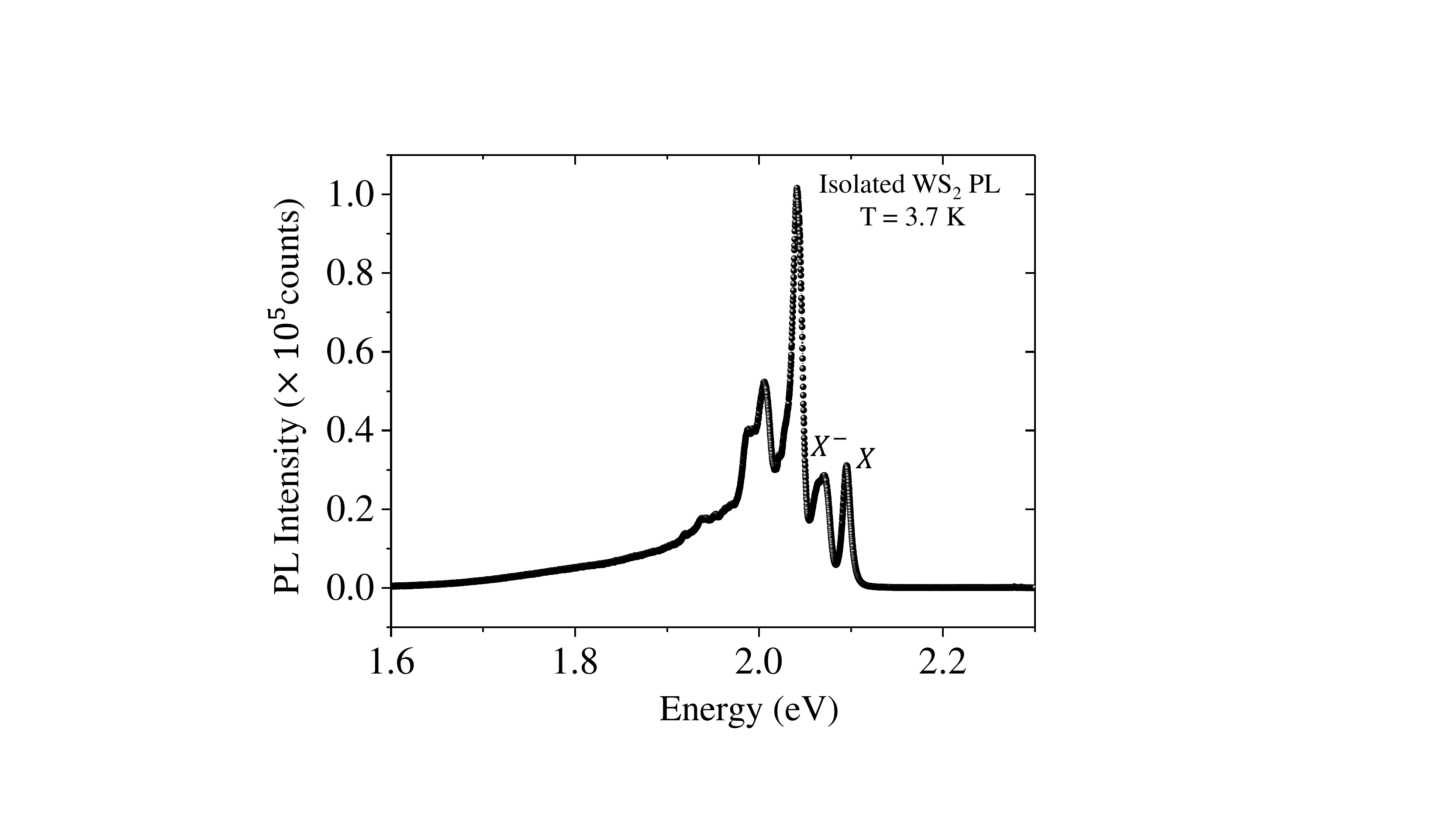}
Figure S3: Low temperature PL intensity profile for monolayer WS\tsub{2} at 3.7 K, indicating sub-bandgap luminescence.
\end{figure}

\newpage
\begin{figure}[!hbt]
\centering
\includegraphics[scale=0.5]{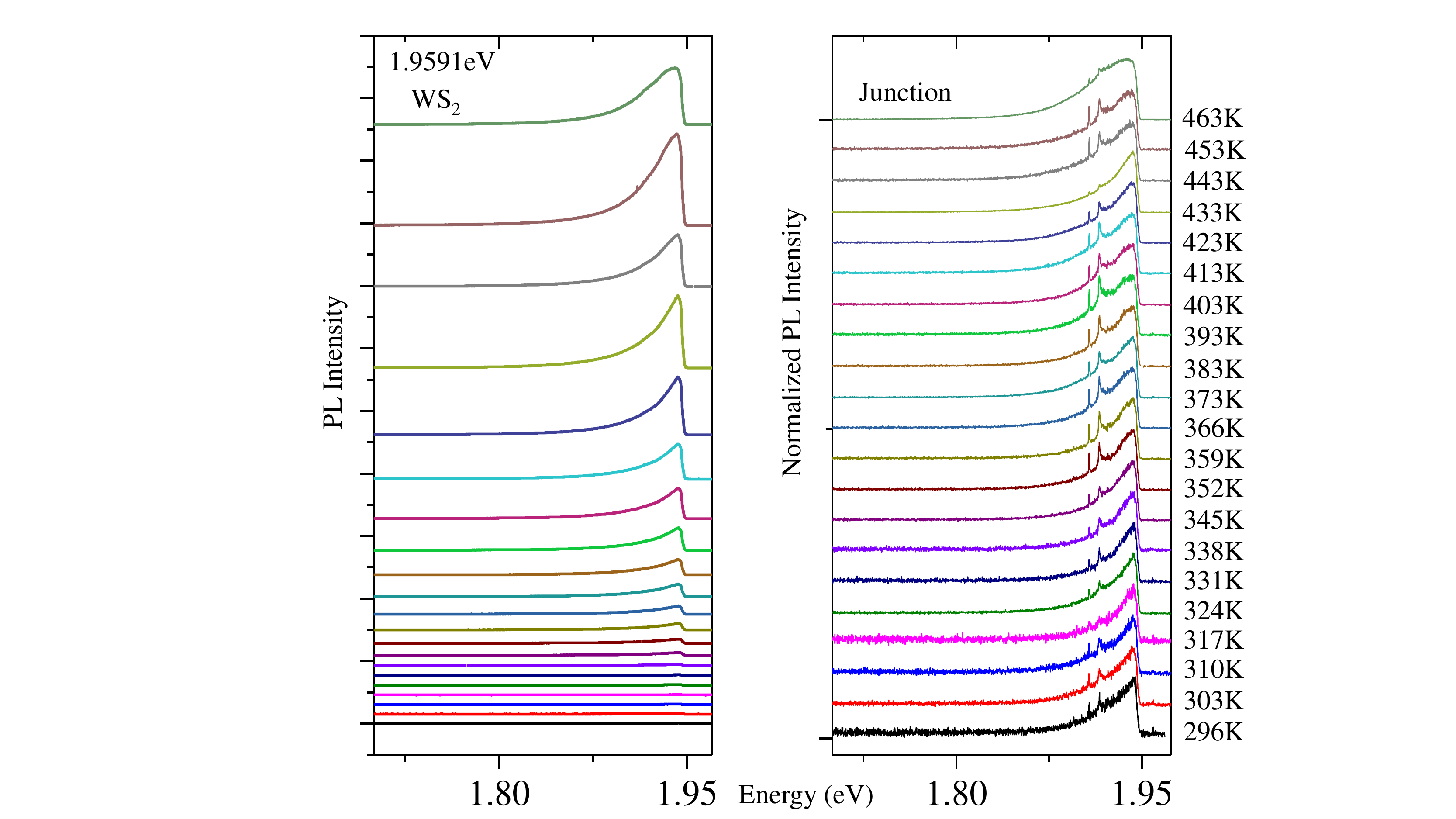}
Figure S4: PL intensity for monolayer WS\tsub2 (left panel) and normalized PL intensity for the junction (right panel) with 1.9591 eV laser excitation. The sharp Raman peaks are observable at the junction due to quenching of background PL.
\end{figure}

\newpage
\begin{figure}[h]
\centering
\includegraphics[scale=0.5]{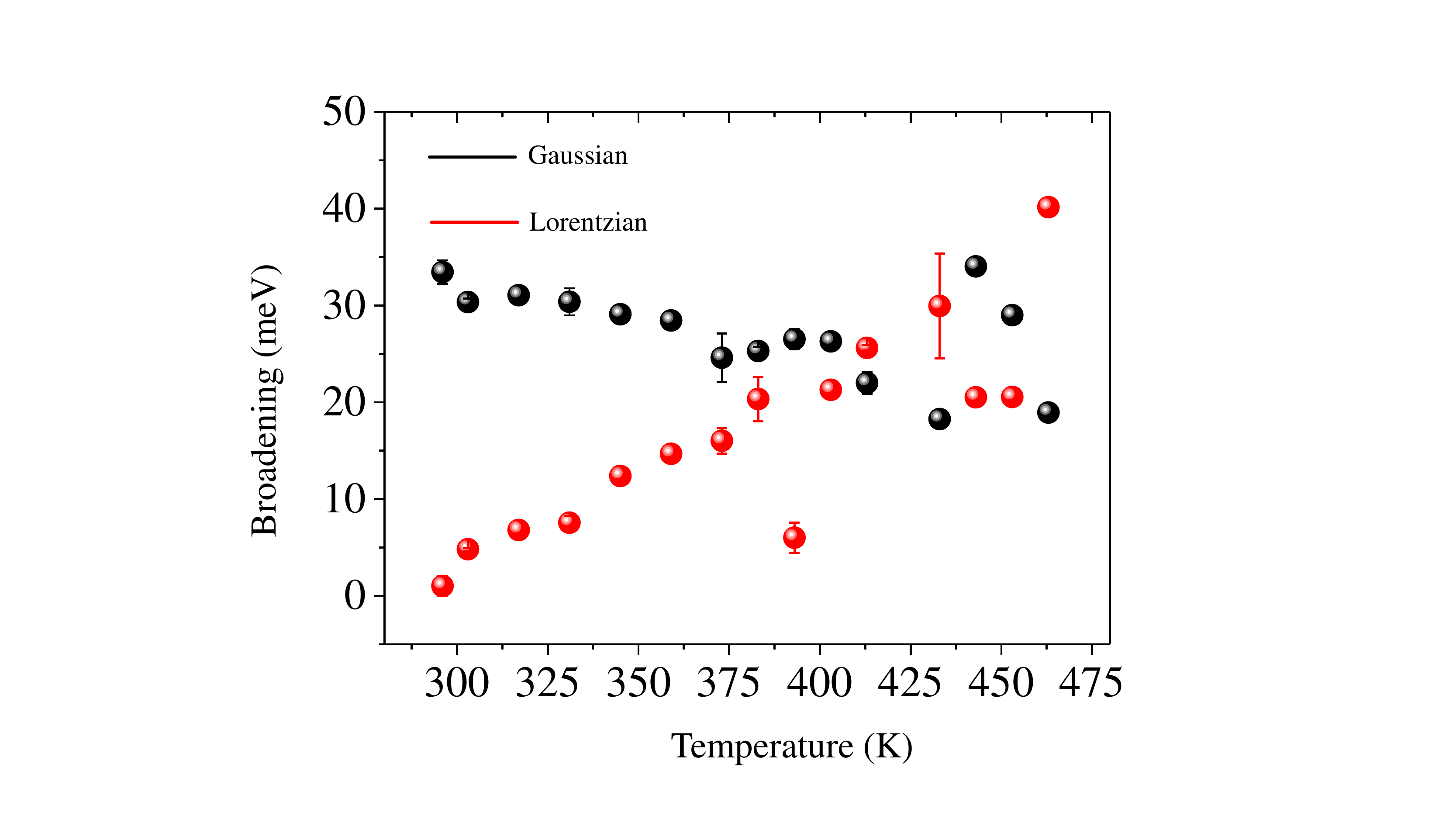}
Figure S5: Extracted Gaussian (inhomogeneous) and Lorentzian (homogeneous) broadening of the exciton at the junction across the temperature range.
\end{figure}

\end{document}